# The protein map of *Synechococcus* sp. PCC 7942 - the first overlook*

Olga A. Koksharova[#‡] ¶, Johan Klint[#], and Ulla Rasmussen[#]

[#] *Department of Botany, Stockholm University, SE-106 91 Stockholm, Sweden*

**Running title:** Proteomics of *Synechococcus* sp. PCC 7942.

¶Corresponding author. Telephone: (7)-095-137-8795 Fax: (007)-095-132-8962.

*E-mail address:* OA-Koksharova@rambler.ru

[‡] Present address: N.I. Vavilov Institute of General Genetics, Gubkin str.,3, 119991 Moscow, Russia

# SUMMARY

The unicellular cyanobacterium *Synechococcus* PCC 7942 has been used as a model organism for studies of prokaryotic circadian rhythms, carbon-concentrating mechanisms, response to a variety of nutrient and environmental stresses, and cell division. This paper presents the results of the first proteomic exploratory study of *Synechococcus* PCC 7942. The proteome was analyzed using two-dimensional gel electrophoresis followed by MALDI-TOF mass spectroscopy, and database searching. Of 140 analyzed protein spots, 110 were successfully identified as 62 different proteins, many of which occurred as multiple spots on the gel. The identified proteins were organized into 18 different functional categories reflecting the major metabolic and cellular processes occurring in the cyanobacterial cells in the exponential growth phase. Among the identified proteins, 14 previously unknown or considered to be hypothetical are here shown to be true gene products in *Synechococcus* sp. PCC 7942, and may be helpful for annotation of the newly sequenced genome.

## The ABBREVIATIONS used are:

ACN, acetonitrile; CHAPS, 3-[(3-cholamidopropyl)dimethyl-amonio]-1-propanesulfonate; CHCA, α-Cyano-4-hydroxycinnamic acid;2-D, two-dimensional; DTT, dithiothreitol; IPG, immobilized pH gradient; kDa, kilodalton (molecular mass); MALDI-TOF, matrix-assisted laser desorption/ionization-time of flight; MS, mass spectrometry; PAGE, polyacrylamide gel electrophoresis; SDS, sodium dodecyl sulfate; TFA, trifluoroacetic acid; TPR, the tetratricopeptide repeat .

# INTRODUCTION

The unicellular *Synechococcus* sp. PCC 7942 is an obligate photoautotroph, belonging to the ancient cyanobacterial group of photoautotrophic prokaryotes (1, 2). *Synechococcus* sp. PCC 7942 has been extensively used as a model organism in biochemical, physiological, and genetic studies of important cellular processes, such as the prokaryotic circadian clock (3), the cyanobacterial carbon-concentrating mechanism (4, 5, 6), nitrate metabolism (7, 8, 9), response to iron deprivation (10, 11, 12) and to a variety of nutrient stresses (13, 14), as well as adaptation to variations in ambient temperature (15, 16) and light intensity (17). With an oxygenic photosynthesis similar to that of green plants, cyanobacteria are believed to be the ancestors of chloroplasts in higher plants as a result of an ancient endosymbiosis (18). Cyanobacteria may therefore be used as model organisms in studying plant photosynthesis, and their potential is being enhanced by means of available genetic tools and genomic sequences (19). In fact, *Synechococcus* sp. PCC 7942 has recently been used in studying the genetic control of cyanobacterial cell division and plastid division in higher plants (20, 21, 22).

No proteomic research has previously been published for *Synechococcus* sp. PCC 7942; however, over the past few years, proteomic data concerning a number of cyanobacteria have been published (23). As the recently sequenced genome was released in publicly accessible databases (http://genome.jgi-

psf.org/finished_microbes/synel/synel.home.html), initiation of proteomic studies in this organism is enabled and highly motivated. The existence of proteins annotated as "hypothetical" can be verified by proteomic methods, and such verification of predicted gene products is an important step in attempting to achieve a correctly annotated genome. Post-translational modification of proteins is often not apparent from DNA sequence analysis, but is usually visible in proteomic work based on two-dimensional (2-D) gel electrophoresis. The level of such modifications can be monitored by identifying protein spots encoded by the same gene but occurring in different positions on the 2-D gels. These modifications might eventually be related to the function or activity of a protein (24).

In this study proteomic work for *Synechococcus* sp. PCC 7942 was initiated and the first protein map of this cyanobacterium is presented. The protein identification was done using MALDI-TOF mass spectroscopy and database searching. The obtained results give insight into the functional and metabolic processes occurring in an important model organism, and may additionally be significant for the annotation and interpretation of the newly sequenced genome of *Synechococcus* sp. PCC 7942.

**EXPERIMENTAL PROCEDURES**

*Cultivation of Synechococcus sp. PCC 7942*-The wild-type strain of *Synechococcus* sp. PCC 7942 was grown in BG11 medium (2) in 100-ml Erlenmeyer flasks at 25ºC in continuous light (18 microEinsteins $m^{-2}s^{-1}$) on a rotary shaker for three days, and harvested when reaching the culture density 0.4-0.5 measured at $OD_{665}$

*Protein Extraction and Separation*-Cultures of 50 ml were collected by centrifugation (3,000 x g) at 4ºC for 5 min, and frozen at -75ºC until use. The cyanobacterial pellets were dissolved in 300 μl of extraction buffer (8M Urea, 0.5% [v/v] IPG buffer [Amersham Biosciences, Sweden] pH 3-10, 2% [w/v] CHAPS, bromophenol blue [traces], protease inhibitors [Protease Inhibitor Cocktail Tablets, Roche Diagnostics Scandinavia AB, Sweden], and 0.28% DTT [w/v]). The samples were then subjected to five cycles of freezing in liquid nitrogen and thawing followed by six cycles of sonication (BANDELIN Sonopuls HD 2070 MS72, DPC Scandinavia AB, Sweden) for 2 seconds each, intervened by cooling for 40 seconds on ice. Cell debris was pelleted by centrifugation using a bench top centrifuge (15,800 x g) for 10 min at room temperature. The supernatants with total proteins were retrieved and stored at -20ºC until use. Protein concentration was measured with the RC DC Protein Assay kit (Bio-Rad, Richmond, CA, USA).

*Two-D Gel Electrophoresis*-Eighteen cm immobiline gel strips (Amersham Biosciences, Sweden) with a pH range 3-10 were rehydrated in extraction buffer (see protein extraction) over night at room temperature. Prior to isoelectrical focusing, 800 μg protein extracts were loaded onto the gel strips in application cups positioned at the acidic end of the strip. The first-dimensional electrophoresis was performed on an IPGphore (Amersham Biosciences, Sweden) with voltage settings of 500 V for 1 h, 1,000 V for 1 h, 8,000 V for approximately 10 h. The Vhrs were monitored, and the last focusing step was interrupted when the run reached a total of 80,000 Vhrs. Before the second dimension SDS-PAGE the IPG strips were equilibrated for 15 min in 50 mM Tris-HCl [pH 8.8], 6 M urea, 30% (v/v) Glycerol [87%], 2% SDS, 10 mg · $ml^{-1}$ DTT followed by 15 min in the same buffer with 25 mg $ml^{-1}$ iodoacetamide instead of DTT. After equilibration the strip was placed on top of a 12% polyacrylamide gel and embedded by the addition of heated 0.5% (w/v) low-melting agarose in SDS electrophoresis running buffer (25 mM Tris, 192 mM glycine, 0.1% SDS). SDS-PAGE was performed in a PROTEAN II® xi cell gel electrophoresis unit (BioRad). After separation (10 mA/gel for 15 min followed by 20 mA/gel until the dye front reached the bottom edge of the gel), gel was incubated in fix solution (20% (v/v) methanol and 10% (v/v) acetic acid) for 4 h and subsequently stained for proteins with Coomassie Brilliant Blue R250 (Fluka Chemie AG, Switzerland) for 24 h. Destaining was accomplished overnight (20% (v/v) methanol and 10% (v/v) acetic acid) at 4ºC, followed by a distilled water wash. Gel imaging was performed on a CanoScan 5000F (Canon, Japan). As protein molecular weight standard BenchMark™ Pre-Stained Protein Ladder (Invitrogen, USA) was used.

*In-Gel Digestion and Peptide Extraction*-Spots were excised from the gel, destained with 20% methanol and 10% acetic acid and in gel digestion with trypsin was performed according to Wilm et al. (25) with the following modifications of the protocol. Gel pieces were washed three times in 200 μl water, followed by two times in 200 μl of 25 mM ammonium bicarbonate in 50% (v/v) ACN. Reduction and alkylation steps were omitted because cysteines were carbamidomethylated during 2-D electrophoresis. Gel pieces were dehydrated with 100% ACN (200 μl). Sequence grade modified trypsin (Sigma, Sweden), at the concentration 12.5 ng μl$^{-1}$ in digestion buffer (50 mM ammonium bicarbonate and 5 mM CaCl$_2$), was then added to the dehydrated gel pieces and was allowed to be absorbed for 45 min on ice. Excess trypsin was removed and replaced with 10 μl of 50 mM ammonium bicarbonate. In-gel digestion of protein was performed over night at 37ºC, followed by three subsequent extractions of peptides with 50% ACN/5% TFA. Combined supernatants were dried in a vacuum centrifuge (DNA-Plus, Heto Lab Equipment, Denmark) for 2-4 hrs. The dried pellets were resuspended in a small volume (5-10 μl) of 0.1% TFA in 50% ACN and stored at -20ºC until analysis.

*MALDI-TOF Mass Spectrometry Analysis and Protein Identification*-Mass spectra were recorded in positive reflection mode by using an Applied Biosystems MALDI-TOF Voyager-DE STR mass spectrometer equipped with a delayed ion extraction technology. α-Cyano-4-hydroxycinnamic acid (CHCA) was used as the matrix. The time of flight was measured using the following parameters: 20 kV accelerating voltage, 65% grid voltage, 0% guide wire voltage, 200 ns delay, low mass gate of 700 Da and acquisition mass range 800-4000 Da. The peptide mass profiles produced by MS were internally calibrated in the MoverZ software (http://www.genomicsolutionscanada.com) using known autolysis peaks from porcine trypsin. After subtraction of known background peaks derived from the matrix, trypsin and traces of keratin, the lists of peptide masses were compared to databases using the Mascot software (http://www.matrixscience.com). The following search parameters were applied for the Mascot searches: NCBInr was used as the protein sequence database; taxonomy was set to "All entries"; a mass tolerance of 30 ppm was applied with one miss cleavage allowed; possible fixed modification was considered to be alkylation of cysteine by carbamidomethylation, and oxidation of methionine. The program FindMod (http://www.expasy.org/tools/findmod/) was used to find modifications for analysis of unmatched peptide masses. For most peptide mass fingerprints, hits with probability based Mowse score > 76 were obtained, which indicates that the hits were significant ($p<0.05$), and that the proteins were successfully identified. For some cases the Mowse score was < 76, which indicates that the protein was not identified with reliability within the level of significance. These protein spots was excluded from the results unless the top hit was from *Synechococcus* sp. PCC 7942, well separated from hits to other proteins, and unmatched peptides could be explained by FindMod program. If these criteria were fulfilled, the protein spots were considered as reliable hits and are presented.

**RESULTS AND DISCUSSION**

A total number of 140 protein spots were apparent on the 2-D Coomassie blue-stained gel (Fig.1), 110 of which were identified as representing 62 different proteins summarized in Table 1. The proteins were categorized in 18 different functional groups (Table 2). This protein map gives a snap-shot of the ongoing metabolic and cellular processes occurring in an exponentially growing culture. Twenty-three proteins were found as sets of several spots. In addition, 14 unknown and hypothetical proteins were localized on the protein map of *Synechococcus* sp. PCC 7942 for the first time. The identified proteins were overrepresented by soluble proteins which were expected due to the protein extraction used, favouring detection of such proteins. Further development and optimization of protein fractionation methods are needed for the visualization of the hydrophobic part of the *Synechococcus* proteome. Some of the identified proteins are described in further detail below.

*Proteins Involved in Cell Morphogenesis (Group 1)*-A set of spots (1; 11a, b, c and 24) represented proteins related to cell division, which reflect the frequent cell divisions occurring in an exponentially growing culture. Spot 1 represented a molecular chaperone comprising a 634-amino-acid protein containing two overlapping regions. The first region (1 to 371 amino acids), named MreB, has been described as an actin-like ATPase involved in cell morphogenesis (cell division and chromosome partitioning) (26). The second region (4 to 597 amino acids), named Hsp70, is an ATP-dependent chaperone that helps in protein folding (27). It is notable that MreB is a shape-determining protein (28), homologous to Hsp70 and to the FtsA protein, which is involved in bacterial cell division (29). Another actin-like ATPase (MreB) was identified as spot 24 (Fig. 1, Table 2, group 1). The cell division protein FtsZ (spots 11 a, b, c) was also detected, occurring as a triplet of spots with slightly different pI values and molecular weights (Fig. 1). This protein is a GTPase involved in septum formation and is considered to be the key proteins during bacterial cell division (30, 31).

*Cell Envelope Biogenesis and Peptidoglycan Synthesis Proteins (Groups 2 and 3*-An actively growing cell culture, such as the one used in this study, require constant new building of structural components in the cell wall. This was reflected in the proteome by the presence of proteins involved in such processes. Spot 5 represents an outer membrane efflux protein possessing a 28-amino-acid signal peptide as predicted by the SignalP software at http://www.cbs.dtu.dk/services/SignalP/ (32). Members of this protein family, the OEP family, form trimeric channels allowing export of a variety of substrates in Gram-negative bacteria (33). Another protein (spots 13 a, b) was identified with 89% sequence identity to the SomA protein from the closely related cyanobacterial strain, *Synechococcus* sp. PCC 6301 (34). Tris protein possesses a 24-amino-acid, cleavable signal peptide involved in protein targeting to the outer membrane (35). SomA is a porin in the outer membrane of this unicellular cyanobacterium. SomA and the homologous protein in spots 13a and b contain S-layer homology (SLH) domain.

A component of the type II secretory pathway, HofQ (spots 14 a, b, c), was found as three separate spots with different pI values. This protein contains one transmembrane helix (residue 266–283 amino acids) as well as an N-terminal, 27-amino-acid signal peptide. The protein is involved in the general (type II) secretion pathway (GSP) in Gram-negative bacteria, a signal sequence-dependent process responsible for protein export (36, 37).

The organic solvent-tolerant protein OstA (spot 46) belongs to a protein family that is mostly uncharacterized. However, in *Escherichia coli*, OstA has been shown to be organic solvent tolerant (38, 39). This protein determines N-hexane tolerance and is involved in outer membrane permeability. The protein is essential for envelope biogenesis and could be part of a targeting system for outer membrane components.

UDP-N-acetylmuramyl tripeptide synthase (spots 7 a, b, c) and diaminopimelate epimerase (spots 33 a, b) are both involved in peptidoglycan synthesis. The former enzyme belongs to the Mur ligase family and possesses a Mur ligase C domain. The later enzyme is involved in the lysine biosynthetic pathway by catalyzing *meso*-diaminopimelate (*meso*-DAP), the direct precursor of lysine and an essential component of cell wall peptidoglycan in Gram-negative bacteria (40).

*Protein Synthesis and Post-Translational Protein Processing (Group 4 and 5*-Two different RNA-binding proteins (spots 53 and 59) containing RRM domain (RNA recognition motif), involved in RNA binding and the regulation of transcription termination, were identified. It has been noted that cyanobacterial RNA-binding proteins (Rbp) and eukaryotic glycine-rich RNA-binding proteins (GRPs) are similar in both structure and regulation, as a result of convergent evolution (41). *Rpb* of cyanobacteria are under stress-responsive regulation and the expression of *rpb* genes in *Anabaena variabilis,* strain M3, is induced by low temperature (as is also the case for GRPs in plants and mammals) (41, 42). Additionally, it has been shown that the transcription of *rbp* genes in *Anabaena* sp. PCC 7120 is also under osmotic stress regulation (43).

Ribosomal protein S1, identified as spot 28, possesses three ribosomal protein S1-like RNA-binding domains. Ribosomal protein S1 is essential for cell viability and is part of the 70S ribosome in bacteria (44). Furthermore, ribosome recycling factor was identified as spot 40; it is a protein that dissociates the ribosome from the mRNA after termination of translation and is essential for bacterial growth. A translation elongation factor, Ts, was identified as two spots (29 a, b) of different pI values. An Asp-tRNAAsn/Glu-tRNAGln amidotransferase B subunit is present on the protein map as two spots (15 a, b), also slightly separated by a difference in pI.

Seven identified proteins are involved in post-translational protein processing and protein folding (Table 2, group 5). From the HSP60 chaperonin family two different chaperonin GroEL (spots 2 a, b and spots 4 a, b, c, d, e, f) proteins were identified. The two proteins show 62% amino acid identity with each other. In addition, a protein was identified as a GroES chaperone (spot 54). The GroEL/GroES system is a major chaperone system in all bacteria, and the mechanisms by which this system responds to stress conditions have been extensively studied in cyanobacteria (45, 46, 47). The fourth identified chaperone is a molecular chaperone GrpE (spot 34). This protein stimulates, jointly with DnaJ, the ATPase activity of the DnaK chaperone in prokaryotes (48). In addition, a periplasmic protease (spots 18 a, b) possessing an N-terminal transmembrane domain was identified.

FKB-type peptidyl-prolyl cis-trans isomerase 1 (spot 61) accelerates protein folding by catalyzing the cis-trans isomerization of proline imidic peptide bonds in oligopeptides. Another isomerase, cyclophilin (peptidylprolyl isomerase or rotamase), was identified in spots 42 a, b. Cyclophilins fold helper enzymes and represent a family within the enzyme class of peptidyl-prolyl cis-trans isomerases (49).

*Protein–Protein Interactions (Group 6)*-Two different proteins (spots 12 and 31) were identified as FOG: TPR repeat-containing proteins. The tetratricopeptide repeat (TPR) is a degenerate 34-amino-acid sequence identified in a wide variety of proteins, present in tandem arrays of 1–16 motifs that form scaffolds mediating protein–protein interactions and often the assembly of multiprotein complexes (50). There is extensive evidence showing that TPR motifs are important for the functioning of chaperone, cell-cycle, transcription, and protein-transport complexes (51). It has recently been found that a TPR repeat is present in the cell division protein FTN2 in *Synechococcus* sp. PCC 7942 (21), and in a protein (gene *sll*0886) required for light-activated heterotrophic growth in *Synechocystis* sp. PCC 6803 (52). In this study we have identified two new TPR repeat-containing proteins in *Synechococcus* sp. PCC 7942, the function of which is unknown.

*Regulatory Function (Group 7)*-A methyl-accepting chemotaxis protein (spot 3), classified as a regulatory protein, homologous to a methyl-accepting chemotaxis protein required for the biogenesis of thick pilli and to a phytochrome-like photoreceptor protein involved in positive phototaxis in *Synechocystis* sp. PCC 6803 (53, 54).

*Proteins of General Cyanobacteria Cell Metabolism (Groups 8 - 16)*-Many proteins were identified as sets of two or more spots on the 2-D gel. Two proteins (Group 14, spots 19 a, b, c, d, e and 58 a, b) are involved in a carbon dioxide-concentrating mechanism and carbon dioxide fixation. A small carboxysome shell protein was identified as a set of five spots, four of which have a predictable molecular weight of approximately 11 kDa (spots 19 b, c, d, e, Fig.1 and Table 1) and differences in pI. Spot 19a has a higher molecular weight (approximately 44 kDa) and may represent a tetramer of this small protein, which has not been dissociated during the electrophoretic separation. Carboxysome shell protein belongs to a family of bacterial microcompartment proteins.

Photosynthesis-related proteins (Group 16) were presented by several proteins, including phycobilisome protein (39 a, b, c, d, e, f, g), plastocyanin (spot 57), and two components of photosystem I, namely PsaD (spot 60) and PsaE (spot 62).

All proteins involved in glucose metabolism (Group 10) occurred as a set of multiple spots on the 2-D map of *Synechococcus*. For example, enolase (spots 9 a, b, c) and fructose-1,6-bisphosphatase/sedoheptulose 1,7-bisphosphatase (spots 21 a, b, c) were both present as triple spots mainly differing in pI. Fructose/tagatose bisphosphate aldolase was also identified in three different spots (spots 22 a, b, c), two of which differed slightly in molecular weight and the third one in pI (Fig. 1).

*Oxidative Stress Defense Proteins (Group 17)*-Being a photosynthetic organism, *Synechococcus* sp. PCC 7942 is likely to encounter the formation of free radicals as a byproduct of the oxygenic photosynthesis. Five proteins possibly involved in oxidative stress defense were identified: a glutathione S-transferase (spot 30) and four different peroxiredoxins (spots 35, 43, 45 and 52 a, b). Peroxiredoxins are ubiquitous thioredoxin- or glutaredoxin-dependent peroxidases, functioning in the disposal of peroxides. The peroxiredoxins identified in this study possess the Ahp/TSA family region. This family contains proteins related to alkyl hydroperoxide reductase (AhpC) and thiol-specific antioxidant (TSA). Anti-oxidative stress systems in cyanobacteria have been studied more extensively in *Synechocystis* sp. PCC 6803 (55, 56, 57).

*Unknown and Hypothetical Proteins (Group 18)*-An unclassified group of proteins (Table 2, group 18) were presented by 14 unknown and hypothetical proteins. One example is the hypothetical protein Selo03002321 (spot 36), which so far only has been found in the genome sequence of *Synechococcus* sp. PCC 7942. An uncharacterized conserved protein (spot 32) possesses a region named "protein of unknown function (DUF541)." Members of this family have only been found in bacteria and in the mouse, where it has been named SIMPL (signaling molecule that associates with mouse pelle-like kinase); their function in bacteria is still unknown.

The hypothetical protein (spot 47) contains N-terminal trans-membrane domain. Another unknown protein (spot 48) possesses two cystathionine beta-synthase (CBS) domains. The closest homologues to the protein in spot 48 were found in *Nostoc punctiforme* PCC 73102 (ZP_00109573.1; COG0517: FOG: CBS domain, 65% sequence identity), in *Trichodesmium erythraeum* IMS101 (ZP_00325310.1; COG0517: FOG: CBS domain, 63% sequence identity), and in *Synechocystis* sp. PCC 6803 (NP_441980.1; IMP dehydrogenase, 58 % sequence identity). Although proteins containing the CBS domain carry out a wide range of functions, the general function of the CBS domain is unknown (58).

The predicted membrane protein (spot 49) contains a region named "protein of unknown function (DUF1269)." This family consists of several bacterial and archaeal proteins of approximately 200 residues in length. The function of this family is unknown. However, the closest homologous proteins were found within the *Synechococcus* genome (ZP_00163791.1; COG4803: Predicted membrane protein, 85% sequence identity), in *Synechocystis* sp. PCC 6803 (hypothetical protein *sll*1106, 46% sequence identity), in *Crocosphaera watsonii* WH 8501 (ZP_00178567.1; COG4803: Predicted membrane protein, 46% sequence identity), and in *Gloeobacter violaceus* PCC 7421 (hypothetical protein *gll*1598, 46% sequence identity).

Two additional unknown proteins (spots 55 and 56) possess some specific regions that could give indications as to possible functions. Spot 55 possesses an FAS1 region (four repeated domains in the fasciclin I family of proteins). It has been suggested that the FAS1 domain represents an ancient cell adhesion domain (59). This protein has been annotated in the *Synechococcus elongatus* PCC 7942 genome project (ZP_00165104) as COG2335: secreted and surface protein containing fasciclin-like repeats. The closest homologous proteins were found in *Gloeobacter violaceus* PCC 7421 (hypothetical protein *glr*1006, 67% sequence identity), in *Nostoc* sp. PCC 7120 (hypothetical protein *alr*1320, 62% sequence identity), in *Anabaena variabilis* ATCC 29413 (COG2335: Secreted and surface protein containing fasciclin-like repeats, 62% sequence identity), in *Synechococcus* sp. PCC 7002 (putative secreted protein MPB70, 61% sequence identity), and in *Synechocystis* sp. PCC 6803 (secreted protein MPB70 *sll*1735, 62% sequence identity).

Unknown protein (spot 56) contains a region named "Thioredoxin." Thioredoxins are small (approximately 12 kDa) redox-active proteins that maintain the reductive intracellular redox potential. In oxygenic photosynthetic cells, thioredoxins were recognized as important regulatory proteins in carbon assimilation (60). Regulatory functions of chloroplast thioredoxins have recently been described (61) using a proteomic approach. Three proteins (spots 47, 50, and 51) were annotated as unknown, and show no obvious homology to any known proteins in the public databases. Future targeted inactivation of gene encoding is unknown, and hypothetical proteins could allow discovery of functions of the novel proteins identified in this study.

In conclusion, we have initiated the proteomic study of *Synechococcus* sp. PCC 7942 and identified 62 different proteins, which were functionally organized into 18 different groups. This data gives an overlook of the major metabolic pathways and cellular processes occurring in an actively growing photosynthetic prokaryote, and forms a strong baseline for future proteomic studies in *Synechococcus* sp. PCC 7942. Twenty-three proteins were found as sets of several spots that could present protein post-translational modifications which might regulate the function or activity of these proteins. Fourteen unknown and hypothetical proteins have been localized on the protein map of *Synechococcus* sp. PCC 7942 for the first time. Such verification of products will be an important and useful step in annotating the genome.

*This work was supported by KVA (The Royal Swedish Academy of Science), by Swedish Institute, and by funding connected with project numbers 03-04-49332 and 03-04-48981 from the Russian Foundation for Basic Research.

**Figure legends**

**Figure 1.** Two-dimensional electrophoresis protein index (Coomassie-stained gel) of total protein of *Synechococcus* sp. PCC 7942. Protein molecular weight standard (*y*-axis) and isoelectric points (*x*-axis) are shown. The identification of each labeled spot is presented in Table 1.

**Table 1.** Proteins identified in *Synechococcus* sp. PCC 7942

| No | Protein identification | Putative Mw/pI | Matches | Mowse score | NCBI Accession no |
|---|---|---|---|---|---|
| 1 | Molecular chaperone | 67.7/4.7 | 10/16 | 137 | 46130530 |
| 2a | Chaperonin GroEL (HSP60 family) | 58.2/4.7 | 8/10 | 124 | 53762838 |
| 2b | Chaperonin GroEL (HSP60 family) | 58.2/4.7 | 7/7 | 119 | 53762838 |
| 3 | Methyl-accepting chemotaxis protein | 45.9/4.4 | 7/8 | 118 | 46129853 |
| 4a | Chaperonin GroEL | 58.0/5.1 | 7/8 | 106 | 46130438 |
| 4b | Chaperonin GroEL | 58.0/5.1 | 8/11 | 115 | 46130438 |
| 4c | Chaperonin GroEL | 58.0/5.1 | 8/10 | 115 | 46130438 |
| 4d | Chaperonin GroEL | 58.0/5.1 | 8/11 | 109 | 46130438 |
| 4e | Chaperonin GroEL | 58.0/5.1 | 11/14 | 160 | 46130438 |
| 4f | Chaperonin GroEL | 58.0/5.1 | 6/8 | 89 | 46130438 |
| 5 | Outer membrane protein | 57.6/5.2 | 9/11 | 137 | 45513235 |
| 6a | Unnamed protein | 52.2/5.0 | 11/12 | 184 | 48024 |
| 6b | Unnamed protein | 52.2/5.0 | 10/13 | 152 | 48024 |
| 6c | Unnamed protein | 52.2/5.0 | 8/11 | 121 | 48024 |
| 6d | Unnamed protein | 52.2/5.0 | 15/17 | 238 | 48024 |
| 7a | UDP-N-acetylmuramyl tripeptide synthase | 53.1/ 5.1 | 4/7 | 57 | 46129915 |
| 7b | UDP-N-acetylmuramyl tripeptide synthase | 53.1/ 5.1 | 3/6 | 39 | 46129915 |
| 7c | UDP-N-acetylmuramyl tripeptide synthase | 53.1/ 5.1 | 5/12 | 56 | 46129915 |
| 8a | S-adenosylmethionine synthetase | 45.4/5.0 | 6/10 | 89 | 46130527 |
| 8b | S-adenosylmethionine synthetase | 45.4/5.0 | 6/12 | 80 | 46130527 |
| 8c | S-adenosylmethionine synthetase | 45.4/5.0 | 6/8 | 93 | 46130527 |
| 9a | Enolase | 45.4/4.7 | 8/11 | 128 | 53762871 |
| 9b | Enolase | 45.4/4.7 | 7/7 | 132 | 53762871 |
| 9c | Enolase | 45.4/4.7 | 4/7 | 58 | 53762871 |
| 10a | DNA polymerase III beta subunit | 40.5/4.9 | 4/8 | 70 | 974615 |
| 10b | DNA polymerase III beta subunit | 40.5/4.9 | 9/10 | 168 | 974615 |
| 11a | Cell division GTPase | 40.2/4.9 | 4/5 | 72 | 46130477 |
| 11b | Cell division GTPase | 40.2/4.9 | 5/9 | 77 | 46130477 |
| 11c | Cell division GTPase | 40.2/4.9 | 4/8 | 60 | 46130477 |

**Table 1.** Continued

| No | Protein identification | Putative Mw/pI | Matches | Mowse score | NCBI Accession no |
|---|---|---|---|---|---|
| 12 | FOG: TPR repeat | 31.7/5.4 | 6/11 | 85 | 46129572 |
| 13a | PPE-repeat proteins | 57.2/5.4 | 6/8 | 85 | 46129905 |
| 13b | PPE-repeat proteins | 57.2/5.4 | 5/5 | 84 | 46129905 |
| 14a | Type II secretory pathway,component HofQ | 80.3/8.4 | 5/6 | 78 | 46130521 |
| 14b | Type II secretory pathway,component HofQ | 80.3/8.4 | 4/8 | 48 | 46130521 |
| 14c | Type II secretory pathway,component HofQ | 80.3/8.4 | 4/7 | 50 | 46130521 |
| 15a | Asp-tRNAAsn/Glu-tRNAGln amidotransferase B subunit (PET112 homolog) | 54.6/5.4 | 5/9 | 62 | 53763184 |
| 15b | Asp-tRNAAsn/Glu-tRNAGln amidotransferase B subunit (PET112 homolog) | 54.6/5.4 | 5/7 | 73 | 53763184 |
| 16a | Sulfite reductase, alpha subunit (flavoprotein) | 44.4/ 5.7 | 3/6 | 44 | 45512552 |
| 16b | Sulfite reductase, alpha subunit (flavoprotein) | 44.4/ 5.7 | 10/17 | 146 | 45512552 |
| 16c | Sulfite reductase, alpha subunit (flavoprotein) | 44.4/ 5.7 | 5/8 | 78 | 45512552 |
| 16d | Sulfite reductase, alpha subunit (flavoprotein) | 44.4/ 5.7 | 7/11 | 106 | 45512552 |
| 16e | Sulfite reductase, alpha subunit (flavoprotein) | 44.4/ 5.7 | 4/5 | 64 | 45512552 |
| 17 | Predicted deinelactone hydrolase | 56.0/6.3 | 8/9 | 134 | 53763021 |
| 18a | Periplasmic protease | 46.2/8.9 | 5/8 | 75 | 53762820 |
| 18b | Periplasmic protease | 46.2/8.9 | 4/7 | 58 | 53762820 |
| 19a | Carbon concentrating mechanism/carboxysome shell protein | 10.9/5.9 | 2/5 | 38 | 46129870 |
| 19b | Carbon dioxide concentrating mechanism/carboxysome shell protein | 10.9/5.9 | 4/7 | 83 | 46129870 |
| 19c | Carbon concentrating mechanism/carboxysome shell protein | 10.9/5.9 | 2/5 | 38 | 46129870 |
| 19d | Carbon concentrating mechanism/carboxysome shell protein | 10.9/5.9 | 3/5 | 64 | 46129870 |

**Table 1.** Continued

| No | Protein identification | Putative Mw/pI | Matc-hes | Mowse score | NCBI Accession no |
|---|---|---|---|---|---|
| 19e | Carbon concentrating mechanism/carboxysome shell protein | 10.9/5.9 | 6/10 | 124 | 46129870 |
| 20a | IMP dehydrogenase/GMP reductase | 40.6/5.3 | 4/9 | 53 | 46130119 |
| 20b | IMP dehydrogenase/GMP reductase | 40.6/5.3 | 5/6 | 86 | 46130119 |
| 20c | IMP dehydrogenase/GMP reductase | 40.6/5.3 | 7/9 | 114 | 46130119 |
| 20d | IMP dehydrogenase/GMP reductase | 40.6/5.3 | 6/7 | 100 | 46130119 |
| 21a | Fructose-1,6-bisphosphatase/sedoheptulose 1,7-bisphosphatase and related proteins | 37.3/5.1 | 5/8 | 70 | 45511855 |
| 21b | Fructose-1,6-bisphosphatase/sedoheptulose 1,7-bisphosphatase and related proteins | 37.3/5.1 | 5/9 | 67 | 45511855 |
| 21c | Fructose-1,6-bisphosphatase/sedoheptulose 1,7-bisphosphatase and related proteins | 37.3/5.1 | 10/13 | 159 | 45511855 |
| 22a | Fructose/tagatose bisphosphate aldolase | 39.1 /5.3 | 4/5 | 70 | 46129889 |
| 22b | Fructose/tagatose bisphosphate aldolase | 39.1 /5.3 | 6/7 | 100 | 46129889 |
| 22c | Fructose/tagatose bisphosphate aldolase | 39.1 /5.3 | 4/7 | 56 | 46129889 |
| 23 | Hypothetical protein | 43.6/6.4 | 8/11 | 129 | 24414820 |
| 24 | Actin-like ATPase involved in cell morphogenesis | 35.0/5.1 | 7/10 | 108 | 53763074 |
| 25 | Hypotheticla protein Selo03001742 | 33.5/4.7 | 4/5 | 67 | 45513861 |
| 26 | Hypothetical protein Selo03000717 | 20.8/5.0 | 4/6 | 76 | 46129880 |
| 27 | Cystein synthase | 34.1/5.5 | 7/9 | 118 | 45512974 |
| 28 | Ribosomal protein S1 | 31.8/5.2 | 5/8 | 80 | 22002498 |
| 29a | Translation elongation factor Ts | 24.4/5.3 | 6/9 | 94 | 46130565 |
| 29b | Translation elongation factor Ts | 24.4/5.3 | 6/7 | 110 | 46130565 |
| 30 | Glutathion S-transferase | 29.1/5.1 | 8/13 | 130 | 45512374 |
| 31 | FOG: TPR repeat | 31.7/5.4 | 6/10 | 98 | 53762940 |
| 32 | Uncharacterized conserved protein | 25.0/6.6 | 5/7 | 92 | 53762884 |
| 33a | Diaminopimelate epimerase | 30.3/4.8 | 6/7 | 110 | 46130191 |

**Table 1.** Continued

| No | Protein identification | Putative Mw/pI | Matches | Mowse score | NCBI Accession no |
|---|---|---|---|---|---|
| 33b | Diaminopimelate epimerase | 30.3/4.8 | 3/5 | 48 | 46130191 |
| 34 | Molecular chaperone GrpE (heat shock protein) | 23.0/4.7 | 6/8 | 108 | 45513516 |
| 35 | Peroxiredoxin | 23.7/4.8 | 7/10 | 128 | 15513856 |
| 36 | Hypothetical protein Selo03002321 | 11.4/4.5 | 3/4 | 68 | 53763101 |
| 37 | Inorganic pyrophosphatase | 19.1/4.6 | 5/9 | 88 | 46129844 |
| 38 | Adenylate kinase and related kinases | 20.3/5.0 | 5/8 | 91 | 45513641 |
| 39a | Unnamed protein product | 18.4/5.2 | 4/7 | 72 | 38900 |
| 39b | Unnamed protein product | 18.4/5.2 | 4/6 | 76 | 38900 |
| 39c | Unnamed protein product | 18.4/5.2 | 5/9 | 89 | 38900 |
| 39d | Unnamed protein product | 18.4/5.2 | 4/7 | 72 | 38900 |
| 39e | Unnamed protein product | 18.4/5.2 | 4/5 | 79 | 38900 |
| 39f | Unnamed protein product | 18.4/5.2 | 3/4 | 60 | 38900 |
| 39g | Unnamed protein product | 18.4/5.2 | 2/4 | 37 | 38900 |
| 40 | Ribosome recycling factor | 19.2/6.4 | 4/6 | 74 | 53762955 |
| 41 | Hypothetical protein | 20.8/5.0 | 5/7 | 97 | 46129880 |
| 42a | Cyclophilin; peptidylprolyl isomerase; rotamase | 15.8/5.7 | 2/2 | 45 | 46489 |
| 42b | Cyclophilin; peptidylprolyl isomerase; rotamase | 15.8/5.7 | 3/4 | 60 | 46489 |
| 43 | Peroxiredoxin | 16.6/5.5 | 5/5 | 118 | 45513396 |
| 44 | Unknown protein | 23.0/5.5 | 3/3 | 67 | 24251256 |
| 45 | Peroxiredoxin | 17.6/5.4 | 3/4 | 58 | 45513279 |
| 46 | Organic solvent tolerance protein OstA | 15.2/9.3 | 3/4 | 61 | 45511847 |
| 47 | Hypothetical protein Selo03002005 | 20.4/8.0 | 4/7 | 78 | 53762906 |
| 48 | Unknown protein | 16.9/5.1 | 3/3 | 70 | 24251259 |
| 49 | Predicted membrane protein | 18.3/4.9 | 3/5 | 57 | 45512224 |
| 50 | Hypothetical protein Selo03000898 | 20.9/6.2 | 3/4 | 64 | 46129994 |

**Table 1.** Continued

| No | Protein identification | Putative Mw/pI | Matches | Mowse score | NCBI Accession no |
|---|---|---|---|---|---|
| 51 | Hypothetical protein Selo03000735 | 16.6/5.1 | 4/5 | 80 | 45512960 |
| 52a | Peroxiredoxin | 15.8/4.5 | 7/9 | 144 | 46130348 |
| 52b | Peroxiredoxin | 15.8/4.5 | 6/7 | 125 | 46130348 |
| 53 | RNA-binding proteins (RRM domain) | 10.7/4.9 | 4/4 | 64 | 45513450 |
| 54 | GroES protein | 10.7/4.7 | 5/7 | 106 | 97572 |
| 55 | Unknown protein | 14.0/4.7 | 3/4 | 64 | 17220757 |
| 56 | Unknown protein | 11.8/4.9 | 2/3 | 44 | 24251253 |
| 57 | Plastocyanin | 13.4/5.5 | 2/3 | 46 | 45512645 |
| 58a | Ribulose bisphosphate carboxylase small subunit | 13.3/5.6 | 3/6 | 55 | 46129874 |
| 58b | Ribulose bisphosphate carboxylase small subunit | 13.3/5.6 | 6/8 | 120 | 46129874 |
| 59 | RNA-binding proteins (RRM domain) | 11.3/5.5 | 4/6 | 79 | 53762887 |
| 60 | psaD protein | 15.6/9.3 | 6/8 | 118 | 226392 |
| 61 | FKB-type peptidyl-prolyl cis-trans isomerases 1 | 18.2/5.3 | 3/6 | 52 | 46130456 |
| 62 | Hypothetical protein Selo03000602 | 8.1/8.0 | 2/2 | 58 | 45512850 |

**Table 2.** Functional groups of proteins identified in *Synechococcus* sp. PCC 7942

| No | Functional group | Protein identification | Spot number |
|---|---|---|---|
| 1 | Cell morphogenesis (Cell division and chromosome partitioning) | Molecular chaperone | 1 |
| | | Cell division GTPase | 11 a,b,c |
| | | Actin-like ATPase involved in cell morphogenesis | 24 |
| 2 | Cell envelope bioenesis, outer membrane / Intracellular trafficking and secretion | Outer membrane efflux protein | 5 |
| | | PPE-repeat proteins | 13 a,b |
| | | Type II secretory pathway, component HofQ | 14 a,b,c |
| | | Organic solvent tolerance protein OstA | 46 |
| 3 | Peptidoglycan synthesis | UDP-N-acetylmuramyl tripeptide synthase | 7 a,b,c |
| | | Diaminopimelate epimerase | 33 a,b |
| 4 | Protein synthesis | Asp-tRNAAsn/Glu-tRNAGln amidotransferase B subunit (PET112 homolog) | 15 a,b |
| | | Ribosomal protein S1 | 28 |
| | | Translation elongation factor Ts | 29 a,b |
| | | Ribosome recycling factor | 40 |
| | | RNA-binding proteins (RRM domain) | 53 |

**Table 2.** Continued

| No | Functional group | Protein identification | Spot number |
|---|---|---|---|
| 4 | Protein synthesis | RNA-binding proteins (RRM domain) | 59 |
| 5 | Posttranslational protein processing, modification, maturation, protein turnover, chaperones | Chaperonin GroEL (HSP60 family) | 2 a,b |
|  |  | Chaperonin GroEL | 4 a,b,c,d,e,f |
|  |  | Periplasmic protease | 18 a,b |
|  |  | Molecular chaperone GrpE (heat shock protein) | 34 |
|  |  | Cyclophilin; peptidylprolyl isomerase; rotamase | 42 a,b |
|  |  | GroES protein | 54 |
|  |  | FKB-type peptidyl-prolyl cis-trans isomerases 1 | 61 |
| 6 | Protein-protein interactions | FOG: TPR repeat | 12 |
|  |  | FOG: TPR repeat | 31 |
| 7 | Regulatory functions | Methyl-accepting chemotaxis protein | 3 |
| 8 | Energy production and conversion | Unnamed protein product ATP synthase alpha/beta family | 6 a,b,c,d |
|  |  | Inorganic pyrophosphatase | 37 |

**Table 2.** Continued

| No | Functional group | Protein identification | Spot number |
|---|---|---|---|
| 8 | Energy production and conversion | Adenylate kinase and related kinases | 38 |
| 9 | Amino-acids metabolism | S-adenosylmethionine synthetase | 8 a,b,c |
|   |   | Cystein synthase | 27 |
| 10 | Glucose metabolism | Enolase | 9 a,b,c |
|   |   | Fructose-1,6-bisphosphatase/sedoheptulose 1,7-bisphosphatase and related proteins | 21 a,b,c |
|   |   | Fructose/tagatose bisphosphate aldolase | 22 a,b,c |
| 11 | DNA synthesis | DNA polymerase III beta subunit | 10 a,b |
| 12 | Sulfate metabolism | Sulfite reductase, alpha subunit (flavoprotein) | 16 a,b, c,d,e |
| 13 | Lipid metabolism | Predicted deinelactone hydrolase | 17 |
| 14 | Carbon dioxide concentrating mechanism and carbon dioxide fixation | Carbon concentrating mechanism/carboxysome shell protein | 19 a,b, c,d,e |
|   |   | Ribulose bisphosphate carboxylase small subunit | 58 a,b |
| 15 | Nucleotide transport and metabolism | IMP dehydrogenase/GMP reductase | 20 a,b, c,d |
| 16 | Photosynthesis | Unnamed protein product; phycobilisome protein | 39 a,b,c,d,e,f,g |
|   |   | Plastocyanin | 57 |

**Table 2.** Continued

| No | Functional group | Protein identification | Spot number |
|---|---|---|---|
| 16 | Photosynthesis | psaD protein, an extrinsic polypeptide of PSI | 60 |
| | | Hypothetical protein Selo03000602; Photosystem I reaction centre subunit IV / PsaE | 62 |
| 17 | Oxidative stress defence | Glutathion S-transferase | 30 |
| | | Peroxiredoxin | 35 |
| | | Peroxiredoxin | 43 |
| | | Peroxiredoxin | 45 |
| | | Peroxiredoxin | 52 a,b |
| 18 | Unknown and hypothetical proteins | Hypothetical protein | 23 |
| | | Hypothetical protein Selo03001742 | 25 |
| | | Hypothetical protein Selo03000717 | 26 |
| | | Uncharacterized conserved protein | 32 |
| | | Hypothetical protein Selo03002321 | 36 |
| | | Hypothetical protein | 41 |
| | | Unknown protein | 44 |
| | | Hypothetical protein Selo03002005 | 47 |
| | | Unknown protein | 48 |

**Table 2.** Continued

| No | Functional group | Protein identification | Spot number |
|---|---|---|---|
| 18 | Unknown and hypothetical proteins | Predicted membrane protein | 49 |
| | | Hypothetical protein Selo03000898 | 50 |
| | | Hypothetical protein Selo03000735 | 51 |
| | | Unknown protein | 55 |
| | | Unknown protein | 56 |

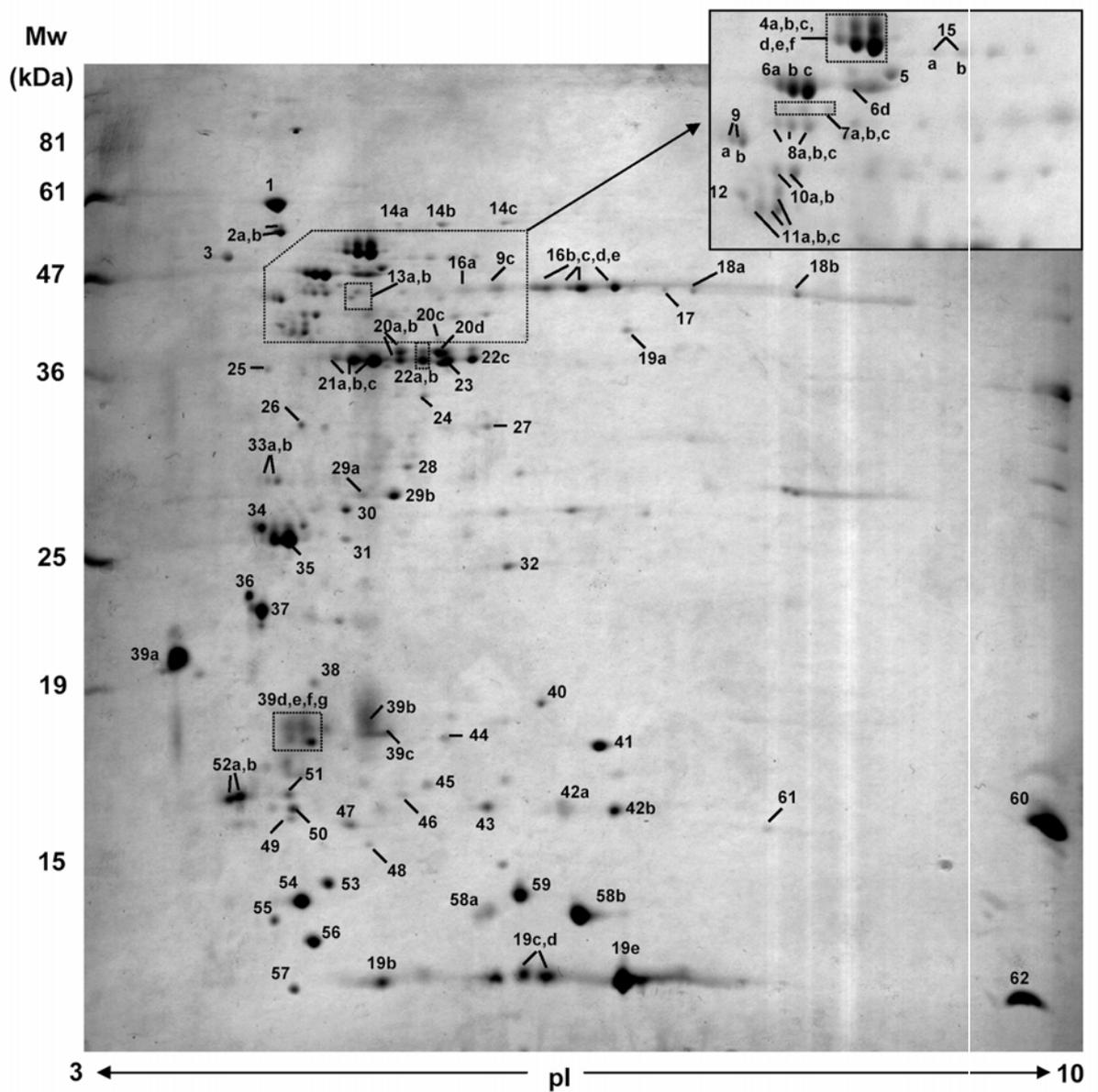

Figure 1. Two-dimensional electrophoresis protein index (Coomassie-stained gel) of total protein of *Synechococcus* sp. PCC 7942.